# Characterization of a graphene-hBN superlattice field effect transistor

Won Beom Choi[1,2], Youngoh Son[1,2], Hangyeol Park[1,2], Yungi Jeong[1,2], Junhyeok Oh[1,2], K. Watanabe[3], T. Taniguchi[4], Joonho Jang[1,2]*

[1] Department of Physics and Astronomy, and Institute of Applied Physics, Seoul National University, Seoul 08826, Korea

[2] Center for Correlated Electron Systems, Institute for Basic Science, Seoul 08826, Korea

[3] Research Center for Electronic and Optical Materials, National Institute for Materials Science, 1-1 Namiki, Tsukuba 305-0044, Japan

[4] Research Center for Materials Nanoarchitectonics, National Institute for Materials Science, 1-1 Namiki, Tsukuba 305-0044, Japan

*Corresponding author's e-mail: joonho.jang@snu.ac.kr

**Abstract**

**Graphene provides a unique platform for hosting high quality 2D electron systems. Encapsulating graphene with hexagonal boron nitride (hBN) to shield it from noisy environments offers the potential to achieve ultrahigh performance nanodevices, such as photodiodes and transistors. However, the absence of a bandgap at the Dirac point presents challenges for using this system as a useful transistor. In this study, we investigated the functionality of hBN-aligned monolayer graphene as a field effect transistor (FET). By precisely aligning the hBN and graphene, bandgaps open at the first Dirac point and at the hole-doped induced Dirac point via an interfacial moiré potential. To characterize this as a submicrometer scale FET, we fabricated a global bottom gate to tune the density of a conducting channel and a local top gate to switch off this channel. This demonstrated that the system could be tuned to an optimal on/off ratio regime by separately controlling the gates. These findings provide a valuable reference point for the further development of FETs based on graphene heterostructures.**

Graphene is excellent platform for studying 2D electron physics due to its high electron mobility and its ability to create diverse heterostructures through stacking and twisting.[1] Graphene is expected to enable ultrahigh performance for applications in electronic and optical devices. However, monolayer graphene, the simplest form, lacks a bandgap, making it difficult to use it to create transistors. Various attempts have been made to build field effect transistors (FETs) using other members of the graphene family. For instance, graphene nanoribbons (GNRs) are quasi one-dimensional structures that exhibit bandgaps in the presence of zigzag and armchair shaped edge morphologies.[2–4] Bernal bilayer graphene has a bandgap when tuned with a vertical electric field, and it has been studied for use in a FET.[5–8] These approaches, however, have their difficulties. GNR devices require precise etching to achieve a thin width dimension of ~10 nm, where the edge disorder significantly reduces mobility to ~1000 $cm^2V^{-1}s^{-1}$, compared to pristine graphene's ~200,000 $cm^2V^{-1}s^{-1}$. Bilayer graphene, despite its high mobility, demands both a vertical electric field for a bandgap and simultaneous gating for density tuning, complicating device design.

In this study, we focus on monolayer graphene, exploring the potential of creating a FET using a heterostructure of monolayer graphene and hexagonal boron nitride (hBN). Theoretically, a bandgap can be generated in monolayer graphene under a periodic electrical potential on the order of several tens of nanometers. However, achieving this with current state-of-the-art nanofabrication processes is challenging.[9–11] An attractive alternative is leveraging a moiré structure. Graphene and hBN are isostructural, with only a 1.8% difference in lattice constants. When an hBN flake and a graphene sheet are aligned with nearly zero angle, the slight lattice mismatch induces a moiré superlattice potential of approximately 12 nm at the interface, opening energy gaps in the monolayer graphene's band structure.[12–15] Previous measurements of this gap-opening effect and the electrical properties of aligned hBN/graphene heterostructures used only a global gate, limiting investigations to bulk properties and precluding the examination of field-effect action of a local gate.[12,16–18] While some studies have used both a local top gate and a global bottom gate[19,20] to investigate electron optics like Fabry-Pèrot resonances, they did not consider the existence of bandgaps, and thus, transistor behaviors were not explored. In this paper, we present aligned hBN/graphene devices for use as FETs, characterizing these devices by fine-tuning their operating conditions to maximize the switching ratio. This shed light on the potential limitations of realizing superlattice-based graphene device.

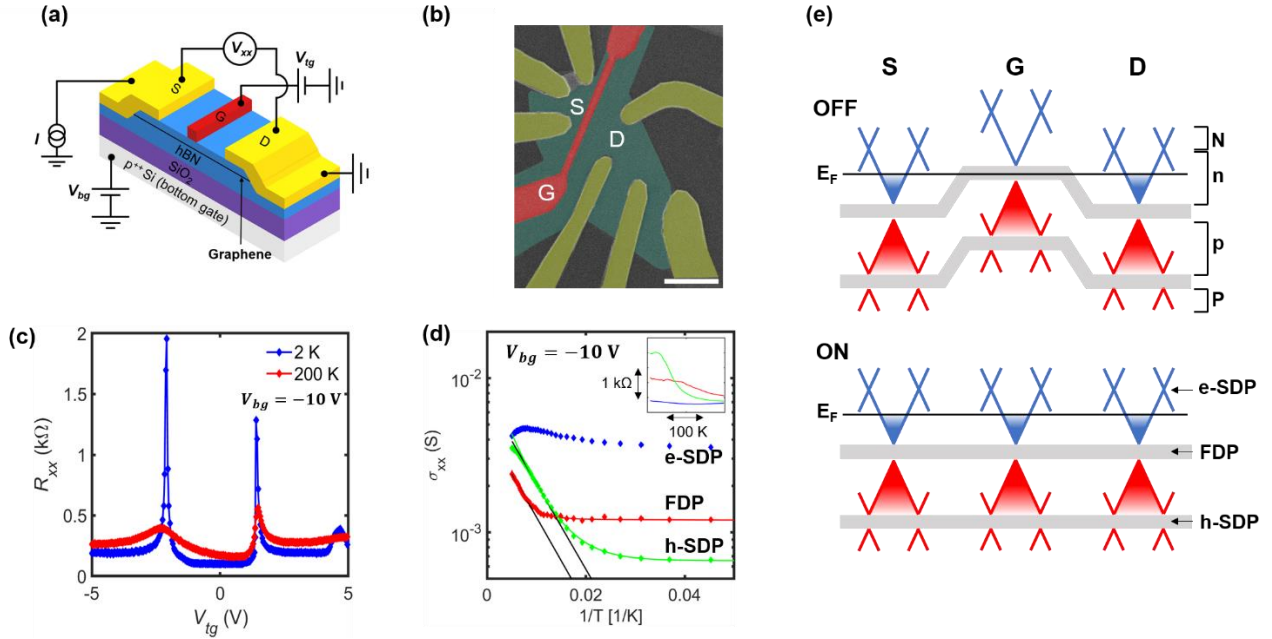

**Fig. 1** (a) Schematic of an hBN/graphene/hBN device with measurement connections. An upper metal layer and an Si-doped layer are used as top and bottom gate, respectively; these can be used to tune the source, drain (S-D) and gate (G) regions, respectively. The thicknesses of the top and bottom hBN layers are 20 and 31 nm. (b) Scanning electron microscope (SEM) image of the device. The scale bar is 3 um, and the width of the top gate, denoted as G, is about 500 nm. (c) Longitudinal resistance $R_{xx}$ as a function of the top gate voltage measured at 2 and 200 K. Three high resistance peaks, corresponding to the h-SDP, FDP, and e-SDP, are clearly visible. (d) Bandgap measurements with a temperature sweep. The gap sizes are obtained from fits to the Arrhenius formula, relating to thermal activation conductance ($\sigma_{TA}$) (black solid line; see text). The green and red solid curves are the fits to the formula combined of a variable-range hopping term and an Arrhenius term, $\sigma = A \cdot exp[-\Delta/(2k_B T)] + B \cdot exp[-(T_h/T)^{1/3}]$. The inset graph is the resistance plot according to temperature. (e) Schematic visualization of 'ON' and 'OFF' cases with spatially gated band diagrams. Note that the device conduction channel consists of three regions in series: one top gated region (region G) and two bottom gated regions (region S-D). The G and S-D regions can independently reach the e-SDP, FDP, and h-SDP.

In **Fig. 1**, we present a representative device [Device A; **Fig. 1(b)**] and the characterization of its electrical properties. In total, we fabricated four distinct hBN-aligned graphene FETs with sufficiently small twist angles to induce bandgaps (see Supplementary Materials).[16–18] The angle alignment and corresponding electrical properties of these devices are highly reproducible with specific fabrication steps. All the devices were fabricated using a conventional dry-transfer method, as follows.[21]

To screen for high-quality hBN pieces, we performed atomic force microscope (AFM)

measurements, selecting samples with a roughness of less than ~200 pm. Each hBN piece was picked up using a polycarbonate (PC)/polydimethylsiloxane (PDMS) stamp under an optical microscope and aligned with a graphene flake after identifying the multiple crystallographic edges of the hBN and the graphene, the angles expected to be multiples of 30°, reflecting the rotational symmetry of the lattice structures.[22] After picking up another hBN piece, now misaligned with the graphene, the completed hBN/graphene/hBN stack was released onto a 300-nm-thick $SiO_2$ wafer. Vacuum annealing was then performed to remove PC residues and small bubbles between the graphene and hBN layers. A second AFM scan determined the position of the graphene and identified a clean area for fabricating a Hall-bar. Finally, e-beam lithography was used to define Cr/Au ohmic contacts and the submicrometer metallic top gate on top of high-quality hBN surfaces as shown in **Fig. 1(b)**.

**Figure. 1(a)** shows a schematic of the setup for four-probe electrical transport measurements of the fabricated device with the top-gated region (G) and bottom-gated regions (S and D). **Figure. 1(c)** shows plots of the device's longitudinal resistance, displaying three peaks due to the modified electronic band structure induced by the moiré potential. The lattice constant of the moiré superlattice potential was determined by identifying the resistance peak for the point of zero carrier density (CNP) at first Dirac point (FDP) and the carrier density of the resistance peak at the second Dirac point (SDP), where the density corresponds to the full filling $4n_0$ – reflecting the spin and valley degrees of freedom of graphene – of the moiré superlattice. The lattice constant of the superlattice was then found from the relationship $1/n_0 = \sqrt{3}\lambda^2/2$. For this device, we found a density of full filling of $2.9 \times 10^{12}$ cm$^{-2}$, corresponding to a lattice constant of 12.7 nm and an hBN-graphene misalignment angle of 0.47°. The Hall mobility of our device was about 100,000 ~ 300,000 cm$^2$V$^{-1}$S$^{-1}$ (**Fig. S4**).

The multi-gate structure of our device allows us to investigate the field-effect action of the 500-nm top gate with high tunability. The Fermi levels in region G and regions S and D were separately controlled by applying two voltage values, $V_{tg}$ and $V_{bg}$, to the top local gate and bottom global gate, respectively. Note that the Fermi levels in regions S and D are identical due to the device geometry. **Figure 1(e)** shows a schematic diagram to visualize the spatially varying profile of band structures and Fermi levels in our devices. To determine the bandgaps within region G [see **Fig. 1(a)**], we measured the top-gate-dependent resistances at various temperatures with a fixed bottom-gate voltage $V_{bg}$ = -10 V corresponding to global p-doping of the device, as shown in **Fig. 1(c)**. The bandgaps (or suppressed density of states) are represented as three resistance peaks: the hole-doped SDP (h-SDP), FDP, and the electron-doped SDP (e-SDP) within region G [see **Fig. 1(e)**]. The resistance peaks at the hole- and electron-doped sides differ in size and width, with bandgaps being much larger on the hole side.[12,14,15] At 200 K, the height of these peaks decreased, indicating insulating states with bandgaps [**Fig. 1(c)**].

For quantitative analysis, we measured resistance as a function of temperature to estimate bandgap sizes of the insulating states. By fitting to the Arrhenius formula, $\sigma \propto exp(-\frac{\Delta}{2k_BT})$, we found bandgaps of approximately 22 and 20 meV at the FDP and h-SDP, respectively. The results are shown in **Fig. 1(d)**. At low temperatures, the conductance is fitted well with the combined formula of an Arrhenius term and a variable-range hopping term,[12,23] as depicted by the red and green curves. For comparison, we performed the same measurements for the entire device, including regions S, D, and G, finding bandgaps of approximately 21 and 34 meV at the FDP and h-SDP, respectively (**Fig. S5**). These values are consistent with previous thermal activation and tunneling measurements using only a global gate.[16–18,24] We found that bandgaps measured using only the top gate in region G were slightly but consistently smaller than those measured using the global bottom gate, attributed to the difference to the charge puddles additionally introduced during the fabrication process of the Ti/Au local top gate, 20 nm away from the graphene layer.[25] Charge puddles can promote thermally activated conduction, whose activation energy is the typical puddle charging energy $E_c$ smaller than the bandgap, and the modified temperature dependent conductivity is given by $\sigma_{TA} = \sigma_0 exp(-E_c/T)$.[26] This explains why measured gap sizes diminish when the region beneath the top metal gate alone is insulating. Otherwise we consider the intrinsic bandgaps in region G to be comparable to those in regions S and D.

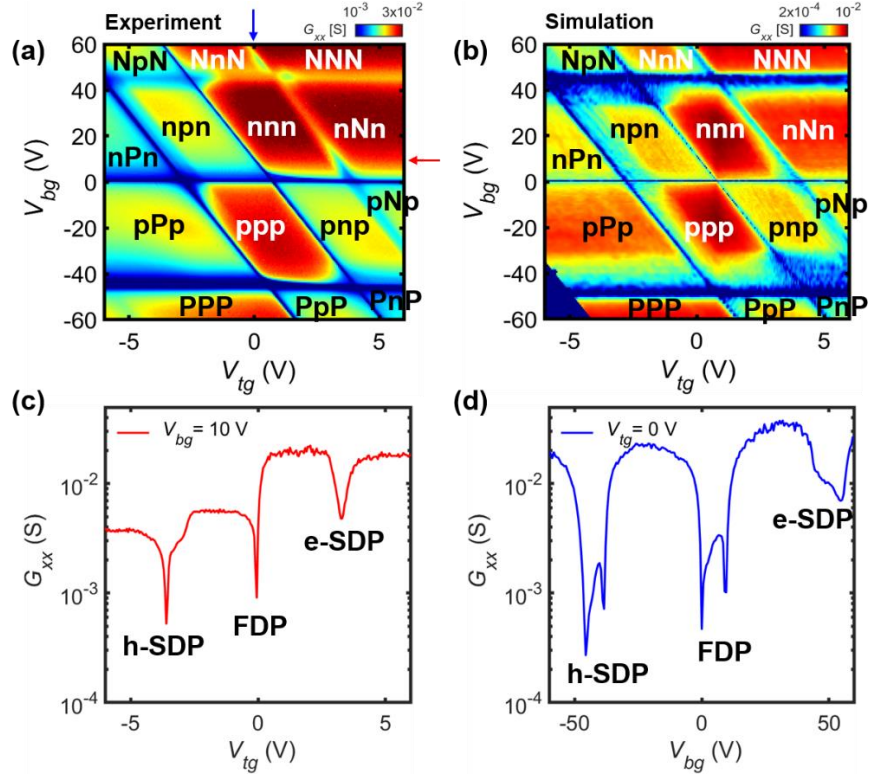

**Fig. 2** (a) Device conductance as a function of $V_{tg}$ and $V_{bg}$ at 2 K. Note that the nnn (ppp) doping regime exhibits higher conductance compared to the npn (pnp) regime. The red and blue arrows refer to the line-cut positions, $V_{bg}=$ 10 V and $V_{tg}=$ 0 V. (b) Corresponding simulation of device conductance as a function of $V_{tg}$ and $V_{bg}$. (c) Conductance vs. $V_{tg}$ measured at $V_{bg}=$ 10 V [see the red arrow in (a)]. The observed dips correspond to three Dirac points; i.e., h-SDP, FDP and e-SDP within region G. (d) Conductance vs. $V_{bg}$ at $V_{tg}=$ 0 V [see the blue arrow in (a)]. Six dips are observed, and these correspond to the Dirac points within regions S-D and G.

Building upon this observation, we further investigated the utility of this locally top-gated device as a FET. Since the Fermi level of region G is affected by both the top and bottom gates, while regions S and D are only affected by the bottom gate, we can systematically study the performance of the field-effect action of region G as a function of the Fermi level of regions S and D by controlling $V_{tg}$ and $V_{bg}$. These features of the device are illustrated in **Figs. 2(a) and 2(b)**, which show how conductance changes with $V_{tg}$ and $V_{bg}$. To investigate the field-effect behavior in detail, we plotted conductance graphs as a function of $V_{tg}$ ($V_{bg}$), at a $V_{bg}=$ 10 V ($V_{tg}=$ 0 V), as shown in **Figs. 2(c) and 2(d)**. These line-cuts, denoted by the arrows in **Fig. 2(a)**, reveal that conductance is substantially reduced at the h-SDP and FDP because region G becomes insulating with ~tens of meV bandgaps, effectively switching-off the conducting channel for the entire device.

In **Fig. 2(d)**, six dips are observed when $V_{bg}$ is varied, implying that, the thin locally top-gated region (G) has superlattice-induced gaps comparable in size to those in the bulk regions (S and D). This confirms that the device performs well, fully developing an insulating state using the local gate in region G.

To study the field-effect action of the top gate in region G as a function of doping within regions S and D, we monitored variations in channel conductance while changing the doping configurations across regions S, D, and G, as shown in **Fig. 2(a)**. We identified the most conductive gate voltage configuration with the highest on/off ratio ($I_{on}/I_{off} = G_{on}/G_{off}$). In **Fig. 2(a)**, we categorize the conductance map of the device in $V_{tg}$ and $V_{bg}$ into 16 panels according to the doping levels of the regions, confirmed by quantum-transport simulations shown in **Fig. 2(b)**. The transport properties were calculated using a tight-binding model with the Kwant software package.[13,27,28] We found that the maximally conducting channel was achieved by tuning both regions S-D and G to doping with carriers of the same-type; to switch off the channel, region G was tuned to one of the insulating states at the FDP and h-SDP while maintaining a high doping level in region S-D.

Interestingly, although the entire device is conductive, the conductance value varies notably depending on the combinations of doping levels in regions G and S-D. When regions S-D and G are doped with the same carrier type (n or p), the Fermi level is positioned in the same band throughout the regions, i.e., valence or conduction band. Conversely, when regions S-D and G are doped with different carrier types, i.e., one region is n (p) and the other is p (n), electrons in the channel encounter a thin insulating region with a bandgap at the FDP somewhere along the channel. This configuration remains conductive, likely due to the existence of hopping-based conductive paths through the thin insulating region in the presence of disorder, but has lower conductance than the nnn or ppp configurations do. Notably, interference effects were observed in the nPn and pPp doping regimes but were absent in other regimes, presumably due to the Berry phase effect in the presence of a bandgap.[29] Therefore, we focus on the device performance in the nnn and ppp doping regimes, which are expected to yield higher on/off ratios than other doping configurations.

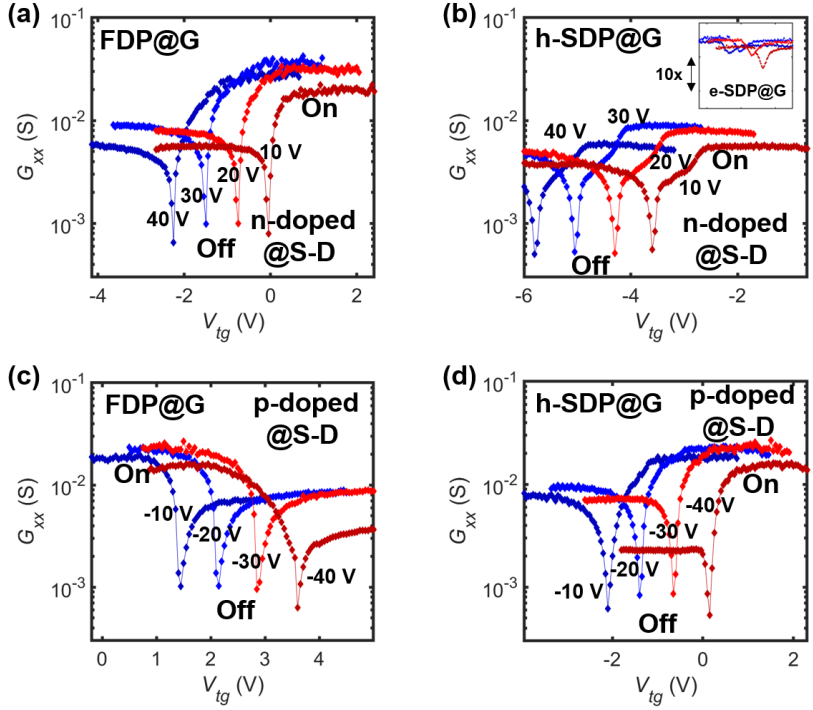

**Fig 3.** (a), (b) Conductance vs. $V_{tg}$ in n-doped@S-D at 2 K. The $V_{bg}$ value was set to 40, 30, 20, and 10 V, from blue to red, to achieve different electron-doping levels. The bandgaps in panels (a) and (b) are at the FDP@G and h-SDP@G. The inset of panel (b) shows the conductance variations at the e-SDP@G. (c), (d) Conductance vs. $V_{tg}$ in p-doped@S-D at 2 K. The $V_{bg}$ value was set to -10, -20, -30, and -40 V, from blue to red, to achieve different hole-doping levels.

To identify the condition exhibiting the highest on/off ratio, a key metric for transistor performance, we performed a quantitative characterization of the device at various device doping levels. **Figure. 3** shows the channel conductances as a function of $V_{tg}$, following the typical gate response curves of an ambipolar transistor. These measurements were taken at various bottom-gate voltages to examine the effects of the doping level in region S-D. As shown in **Figs. 3(a)** and **3(b),** the channel conductance curves undergo interesting changes in shape as the doping level within region S-D varies, particularly in the insulating positions near h-SDP@G and FDP@G. The highest on/off ratio, 45.6, was observed in the 'nnn' regime at $V_{bg}$= 40 V, which also exhibited one of the highest channel conductances. This on/off ratio is compared with those of other high-mobility graphene devices in the Supplementary Material. Given the ambipolar nature of the curves, two on/off ratios were obtained; the larger value for each curve was extracted, and the results are presented in **Table I**. As depicted in the **inset of Fig. 3(b)**, the on/off ratio near e-

SDP@G is an order of magnitude lower than at other positions, suggesting the absence of a bandgap at the e-SDP, as has been theoretically proposed.[15] In the case of p-doping@S-D [**Figs. 3(c)** and **3(d)**], the on/off ratios are smaller than those of the n-doping cases [**Figs. 3(a)** and 3(**b**)]. We attribute the apparent asymmetry between the ppp and nnn conditions to the particle-hole asymmetric effective band mass of the hBN-aligned monolayer graphene system. Based on these observations, we found that the doping control of the source-drain channel is critical to achieving preferred operational conditions, such as high conductance channels and high on/off ratios, for hBN/monolayer graphene transistors and, likely, for future transistors using moiré-induced band structures in general.

**TABLE I.** On/off ratios at various values of the bottom-gate voltage $V_{bg}$. The positive and negative bottom-gate voltages correspond, respectively, to the n-doped@S-D and p-doped@S-D. Each row corresponds to a certain bandgap in region G. The on/off ratios of e-SDP@G could not be measured at p-doped@S-D due to the limit of the top gate voltage.

| On/off ratio | n-doped@S-D | | | | |
|---|---|---|---|---|---|
| Bottom gate | 43 V | 40 V | 30 V | 20 V | 10 V |
| h-SDP@G | 7.9 | 12 | 16.8 | 16.22 | 10 |
| FDP@G | 31 | 45.6 | 37.6 | 31.2 | 25.4 |
| e-SDP@G | 2 | 2.2 | 2.5 | 2.8 | 4.1 |
| On/off ratio | p-doped@S-D | | | | |
| Bottom gate | -10 V | -20 V | -30 V | -40 V | -42 V |
| h-SDP@G | 28.6 | 27 | 26.9 | 29.9 | 25.5 |
| FDP@G | 17.5 | 22 | 23.1 | 25.5 | 17.1 |
| e-SDP@G | 2.3 | 2.4 | – | – | – |

The dual-gate design –with a local gate acting as a switching gate and a global gate controlling the doping of the source-drain channel– appears to be a critical factor for the successful function of the device as a FET. Calculations based on the continuum limit[15] showed that the band structure the near the CNP is asymmetric – i.e. the conduction band and the valence band have quite different dispersions and effective masses – thus, finding an optimal operation point for a FET based on the modified dispersion is vital. Our results demonstrate this asymmetry in the FET performance between the n and p-doped channels (**Table I)**, and the optimized working conditions found in one of the n-doped channels suggest that this prototype FET with a submicron-sized gate behaves as expected, reflecting the theoretically proposed superlattice-modified band structure.

In conclusion, we characterized an aligned hBN/graphene device by varying local top-gate and global bottom-gate voltages. This tunability allowed us to identify an optimal operational condition as a FET. The on/off ratios measured for this device suggest that the design has room for improvement to compete directly with other TMD-based devices. [30,31] This design could also be useful in applications requiring very high mobility, such as quantum devices, optical elements and high-frequency circuits. Our study provides valuable data points from moiré-based FETs with a submicron top gate, supporting the potential utility of moiré-based graphene heterostructures as future FETs by further increasing the bandgaps using techniques such as doubly-aligned hBN-monolayer graphenes and, hBN-encapsulated twisted multilayer graphenes.[32–34]

## Supplementary material

See the supplementary material for additional information about the fabrication process and characterize the device properties.

## Acknowledgments

The work at SNU was supported by the National Research Foundation of Korea grants funded by the Ministry of Science and ICT (Grant Nos. 2019R1C1C1006520, 2020R1A5A1016518, RS-2023-00258359), the Institute for Basic Science of Korea (Grant No. IBS-R009- D1), SNU Core Center for Physical Property Measurements at Extreme Physical Conditions (Grant No. 2021R1A6C101B418), Creative-Pioneering Researcher Program through Seoul National University and Samsung Electronics. J.J. K.W. and T.T. acknowledge support from the JSPS KAKENHI (Grant Numbers 21H05233 and 23H02052) and World Premier International Research Center Initiative (WPI), MEXT, Japan.

## Author declarations

**Conflict of interest**

The authors declare no competing interests.

**Author contributions**

**Won Beom Choi**: Conceptualization (equal); data curation (equal); formal analysis (equal); investigation (equal); Methodology (equal); resources (equal); software (equal); validation (equal); visualization (equal); writing – original draft (equal); writing – review & editing (equal). **Youngoh Son**: Formal analysis (supporting); investigation (supporting); writing – original draft (supporting). **Hangyeol Park**: Formal analysis (supporting); investigation (supporting); writing – original draft (supporting). **Yungi Jeong**: Formal analysis (supporting); investigation (supporting); writing – original draft (supporting). **Junhyeok Oh**: Formal analysis (supporting); investigation (supporting); writing – original draft (supporting). **K. Watanabe**: investigation (supporting) **T. Taniguchi**: investigation (supporting) **Joonho Jang**: Conceptualization (equal); data curation (equal); formal analysis (equal); funding acquisition (lead); investigation (equal); methodology (equal); project administration (lead); resources (equal); software (equal); supervision (lead); validation (equal); visualization (equal); writing – original draft (equal); Writing – review & editing (equal).

**Data availability**

The data that support the findings of this study are available from the corresponding authors upon reasonable request.

**References**

[1] Y. Zhang, Y.-W. Tan, H.L. Stormer, and P. Kim, "Experimental observation of the quantum Hall effect and Berry's phase in graphene," Nature **438**(7065), 201–204 (2005).
[2] M.Y. Han, B. Özyilmaz, Y. Zhang, and P. Kim, "Energy Band-Gap Engineering of Graphene Nanoribbons," Phys. Rev. Lett. **98**(20), 206805 (2007).
[3] X. Li, X. Wang, L. Zhang, S. Lee, and H. Dai, "Chemically Derived, Ultrasmooth Graphene Nanoribbon Semiconductors," Science **319**(5867), 1229–1232 (2008).

[4] L. Liao, J. Bai, R. Cheng, Y.-C. Lin, S. Jiang, Y. Huang, and X. Duan, "Top-Gated Graphene Nanoribbon Transistors with Ultrathin High-k Dielectrics," Nano Lett. **10**(5), 1917–1921 (2010).

[5] B.N. Szafranek, G. Fiori, D. Schall, D. Neumaier, and H. Kurz, "Current Saturation and Voltage Gain in Bilayer Graphene Field Effect Transistors," Nano Lett. **12**(3), 1324–1328 (2012).

[6] A. Varlet, M.-H. Liu, V. Krueckl, D. Bischoff, P. Simonet, K. Watanabe, T. Taniguchi, K. Richter, K. Ensslin, and T. Ihn, "Fabry-P\'erot Interference in Gapped Bilayer Graphene with Broken Anti-Klein Tunneling," Phys. Rev. Lett. **113**(11), 116601 (2014).

[7] T. Uwanno, T. Taniguchi, K. Watanabe, and K. Nagashio, "Electrically Inert h-BN/Bilayer Graphene Interface in All-Two-Dimensional Heterostructure Field Effect Transistors," ACS Appl. Mater. Interfaces **10**(34), 28780–28788 (2018).

[8] R. Du, M.-H. Liu, J. Mohrmann, F. Wu, R. Krupke, H. von Löhneysen, K. Richter, and R. Danneau, "Tuning Anti-Klein to Klein Tunneling in Bilayer Graphene," Phys. Rev. Lett. **121**(12), 127706 (2018).

[9] M. Drienovsky, J. Joachimsmeyer, A. Sandner, M.-H. Liu, T. Taniguchi, K. Watanabe, K. Richter, D. Weiss, and J. Eroms, "Commensurability Oscillations in One-Dimensional Graphene Superlattices," Phys. Rev. Lett. **121**(2), 026806 (2018).

[10] Y. Li, S. Dietrich, C. Forsythe, T. Taniguchi, K. Watanabe, P. Moon, and C.R. Dean, "Anisotropic band flattening in graphene with one-dimensional superlattices," Nature Nanotechnology **16**(5), 525–530 (2021).

[11] S. Berrada, V. Hung Nguyen, D. Querlioz, J. Saint-Martin, A. Alarcón, C. Chassat, A. Bournel, and P. Dollfus, "Graphene nanomesh transistor with high on/off ratio and good saturation behavior," Applied Physics Letters **103**(18), 183509 (2013).

[12] B. Hunt, J.D. Sanchez-Yamagishi, A.F. Young, M. Yankowitz, B.J. LeRoy, K. Watanabe, T. Taniguchi, P. Moon, M. Koshino, P. Jarillo-Herrero, and R.C. Ashoori, "Massive Dirac Fermions and Hofstadter Butterfly in a van der Waals Heterostructure," Science **340**(6139), 1427–1430 (2013).

[13] J.R. Wallbank, A.A. Patel, M. Mucha-Kruczyński, A.K. Geim, and V.I. Fal'ko, "Generic miniband structure of graphene on a hexagonal substrate," Phys. Rev. B **87**(24), 245408 (2013).

[14] L.A. Ponomarenko, R.V. Gorbachev, G.L. Yu, D.C. Elias, R. Jalil, A.A. Patel, A. Mishchenko, A.S. Mayorov, C.R. Woods, J.R. Wallbank, M. Mucha-Kruczynski, B.A. Piot, M. Potemski, I.V. Grigorieva, K.S. Novoselov, F. Guinea, V.I. Fal'ko, and A.K. Geim, "Cloning of Dirac fermions in graphene superlattices," Nature **497**(7451), 594–597 (2013).

[15] P. Moon, and M. Koshino, "Electronic properties of graphene/hexagonal-boron-nitride moiré superlattice," Phys. Rev. B **90**(15), 155406 (2014).

[16] L. Wang, Y. Gao, B. Wen, Z. Han, T. Taniguchi, K. Watanabe, M. Koshino, J. Hone, and C.R. Dean, "Evidence for a fractional fractal quantum Hall effect in graphene superlattices," Science **350**(6265), 1231–1234 (2015).

[17] R. Ribeiro-Palau, C. Zhang, K. Watanabe, T. Taniguchi, J. Hone, and C.R. Dean, "Twistable electronics with dynamically rotatable heterostructures," Science **361**(6403), 690–693 (2018).

[18] H. Kim, N. Leconte, B.L. Chittari, K. Watanabe, T. Taniguchi, A.H. MacDonald, J. Jung, and S. Jung, "Accurate Gap Determination in Monolayer and Bilayer Graphene/h-BN Moiré Superlattices," Nano Lett. **18**(12), 7732–7741 (2018).

[19] C. Handschin, P. Makk, P. Rickhaus, M.-H. Liu, K. Watanabe, T. Taniguchi, K. Richter, and C. Schönenberger, "Fabry-Pérot Resonances in a Graphene/hBN Moiré Superlattice," Nano Lett. **17**(1), 328–333 (2017).

[20] R. Kraft, M.-H. Liu, P.B. Selvasundaram, S.-C. Chen, R. Krupke, K. Richter, and R. Danneau, "Anomalous Cyclotron Motion in Graphene Superlattice Cavities," Phys. Rev. Lett. **125**(21), 217701 (2020).

[21] S. Fan, Q.A. Vu, M.D. Tran, S. Adhikari, and Y.H. Lee, "Transfer assembly for two-dimensional van der Waals heterostructures," 2D Mater. **7**(2), 022005 (2020).

[22] J. Hu, J. Tan, M.M. Al Ezzi, U. Chattopadhyay, J. Gou, Y. Zheng, Z. Wang, J. Chen, R. Thottathil, J. Luo, K. Watanabe, T. Taniguchi, A.T.S. Wee, S. Adam, and A. Ariando, "Controlled alignment of supermoiré lattice in double-aligned graphene heterostructures," Nat Commun **14**(1), 4142 (2023).

[23] G.-H. Lee, D. Jeong, K.-S. Park, Y. Meir, M.-C. Cha, and H.-J. Lee, "Continuous and reversible tuning of the disorder-driven superconductor–insulator transition in bilayer graphene," Sci Rep **5**(1), 13466 (2015).

[24] K. Komatsu, Y. Morita, E. Watanabe, D. Tsuya, K. Watanabe, T. Taniguchi, and S. Moriyama, "Observation of the quantum valley Hall state in ballistic graphene superlattices," Science Advances **4**(5), eaaq0194 (2018).

[25] E. Icking, L. Banszerus, F. Wörtche, F. Volmer, P. Schmidt, C. Steiner, S. Engels, J. Hesselmann, M. Goldsche, K. Watanabe, T. Taniguchi, C. Volk, B. Beschoten, and C. Stampfer, "Transport Spectroscopy of Ultraclean Tunable Band Gaps in Bilayer Graphene," Advanced Electronic Materials **8**(11), 2200510 (2022).

[26] Y. Huang, Y. He, B. Skinner, and B.I. Shklovskii, "Conductivity of two-dimensional narrow gap semiconductors subjected to strong Coulomb disorder," Phys. Rev. B **105**(5), 054206 (2022).

[27] C.W. Groth, M. Wimmer, A.R. Akhmerov, and X. Waintal, "Kwant: a software package for quantum transport," New J. Phys. **16**(6), 063065 (2014).

[28] J. Hu, A.F. Rigosi, J.U. Lee, H.-Y. Lee, Y. Yang, C.-I. Liu, R.E. Elmquist, and D.B. Newell, "Quantum transport in graphene p − n junctions with moiré superlattice modulation," Phys. Rev. B **98**(4), 045412 (2018).

[29] A. Varlet, M.-H. Liu, D. Bischoff, P. Simonet, T. Taniguchi, K. Watanabe, K. Richter, T. Ihn, and K. Ensslin, "Band gap and broken chirality in single-layer and bilayer graphene," Physica Status Solidi (RRL) – Rapid Research Letters **10**(1), 46–57 (2016).

[30] B. Radisavljevic, A. Radenovic, J. Brivio, V. Giacometti, and A. Kis, "Single-layer MoS2 transistors," Nature Nanotech **6**(3), 147–150 (2011).

[31] A. Sebastian, R. Pendurthi, T.H. Choudhury, J.M. Redwing, and S. Das, "Benchmarking monolayer MoS2 and WS2 field-effect transistors," Nat Commun **12**(1), 693 (2021).

[32] L. Wang, S. Zihlmann, M.-H. Liu, P. Makk, K. Watanabe, T. Taniguchi, A. Baumgartner, and C. Schönenberger, "New Generation of Moiré Superlattices in Doubly Aligned hBN/Graphene/hBN Heterostructures," Nano Lett. **19**(4), 2371–2376 (2019).

[33] Z. Wang, Y.B. Wang, J. Yin, E. Tóvári, Y. Yang, L. Lin, M. Holwill, J. Birkbeck, D.J. Perello, S. Xu, J. Zultak, R.V. Gorbachev, A.V. Kretinin, T. Taniguchi, K. Watanabe, S.V. Morozov, M. Anđelković, S.P. Milovanović, L. Covaci, F.M. Peeters, A. Mishchenko, A.K. Geim, K.S. Novoselov, V.I. Fal'ko, A. Knothe, and C.R. Woods, "Composite super-moiré lattices in double-aligned graphene heterostructures," Science Advances **5**(12), eaay8897 (2019).

[34] H. Oka, and M. Koshino, "Fractal energy gaps and topological invariants in hBN/graphene/hBN double moiré systems," Phys. Rev. B **104**(3), 035306 (2021).

# Supplementary Information for "Characterization of the field effect transistor in graphene-hBN superlattice"

Won Beom Choi[1,2], Youngoh Son[1,2], Hangyeol Park[1,2], Yungi Jeong[1,2], Junhyeok Oh[1,2], K. Watanabe[3], T. Taniguchi[4], Joonho Jang[1,2]*

[1] Department of Physics and Astronomy, and Institute of Applied Physics, Seoul National University, Seoul 08826, Korea

[2] Center for Correlated Electron Systems, Institute for Basic Science, Seoul 08826, Korea

[3] Research Center for Electronic and Optical Materials, National Institute for Materials Science, 1-1 Namiki, Tsukuba 305-0044, Japan

[4] Research Center for Materials Nanoarchitectonics, National Institute for Materials Science, 1-1 Namiki, Tsukuba 305-0044, Japan

*Corresponding author's e-mail: joonho.jang@snu.ac.kr

## I. Fabrication method

High-quality hBN and graphene were exfoliated onto a 300 nm thick silicon dioxide ($SiO_2$) layer. A monolayer graphene and ~20 nm hBN were obtained by using color contrast changes indicative of thickness variation on the 300 nm $SiO_2$ substrate. Atomic force microscope (AFM) measurements were conducted after exfoliation to avoid step-like features, hBN layer transitions, and disorder, with hBN exhibiting about 200pm roughness. The presence of layered steps and disorder in hBN induces a local potential and occurs in an inhomogeneous angle between hBN and graphene. Employing a larger hBN compared to graphene ensures complete coverage of graphene regions, facilitating the formation of hBN-only areas and defining the top gate (TG) position without direct contact with graphene. A hBN/graphene/hBN stack was assembled using polycarbonate (PC) and polydimenthylsil-oxane (PDMS) blocks with 100°C. A graphene is aligned with the hBN layer as possible during the pick-up process. Subsequently, this device was transferred onto a Si/SiO$_2$ wafer by increasing the temperature of PDMS block 190 °C to drop down all structure. Removal of the PC film was achieved using N-Methyl-2-pyrrolidone (NMP) followed by cleaning with acetone and IPA. A vacuum annealing process was then conducted at 350°C 5 hours and 500°C 3 hours to eliminate residual contaminants from PC film.[35] As shown in **Fig. S1(a),** the graphene is fully encapsulated by top and bottom hBN layers.

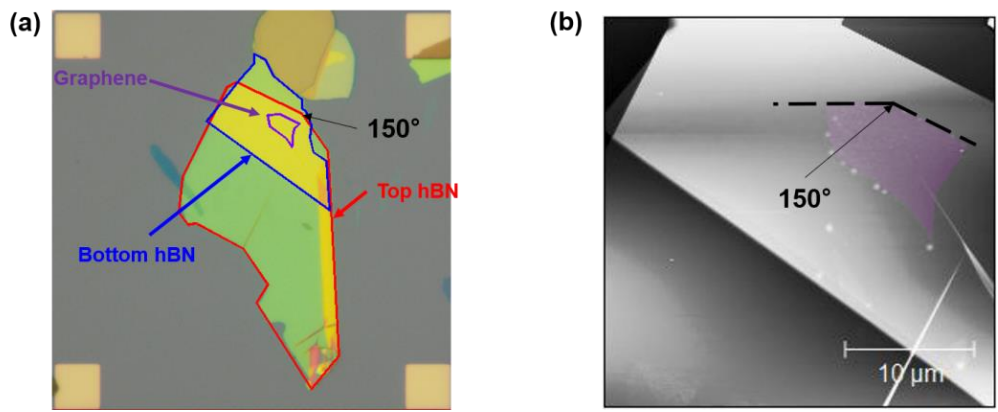

**Fig. S1** (a) The hBN/graphene/hBN structure after the stacking process. The crystal axis of hBN is measured with an OM image. The angle between two different crystal edges should show a multiple of 30°. (b) AFM image of hBN/graphene/hBN (purple). The angle of the graphene edge is about 150°.

The AFM measurement was conducted after this procedure to confirm a sample quality and ensure a clean area to make the device. Electrode and top gate positions were defined with AFM results, followed by the conventional e-beam lithography process. A cold development process was employed for defining the small width of the top gate.

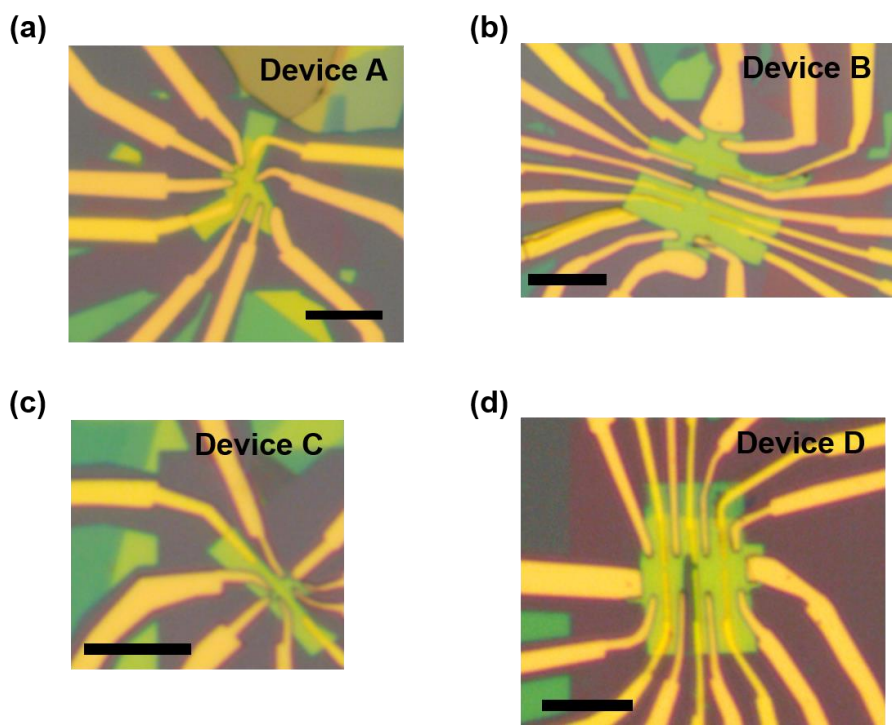

**Fig. S2** (a), (b), (c) and (d) are the hBN aligned graphene heterostructures. Scale bar is 10 um

Ti/Au deposition via e-beam evaporation (5/100 nm) was conducted subsequently. For electrode definition, a one-dimensional edge contact method was employed.[36] Edge parts were defined through RIE utilizing a CF4 (40 sccm), O2 (4 sccm) mixture. Additionally, the sample underwent deposition with a Cr/Au (5/50 nm) layer using e-beam evaporation. Through this sequence of procedures, four different devices, namely device A (main text), device B, device C and device D were fabricated as shown in **Fig. S2**.

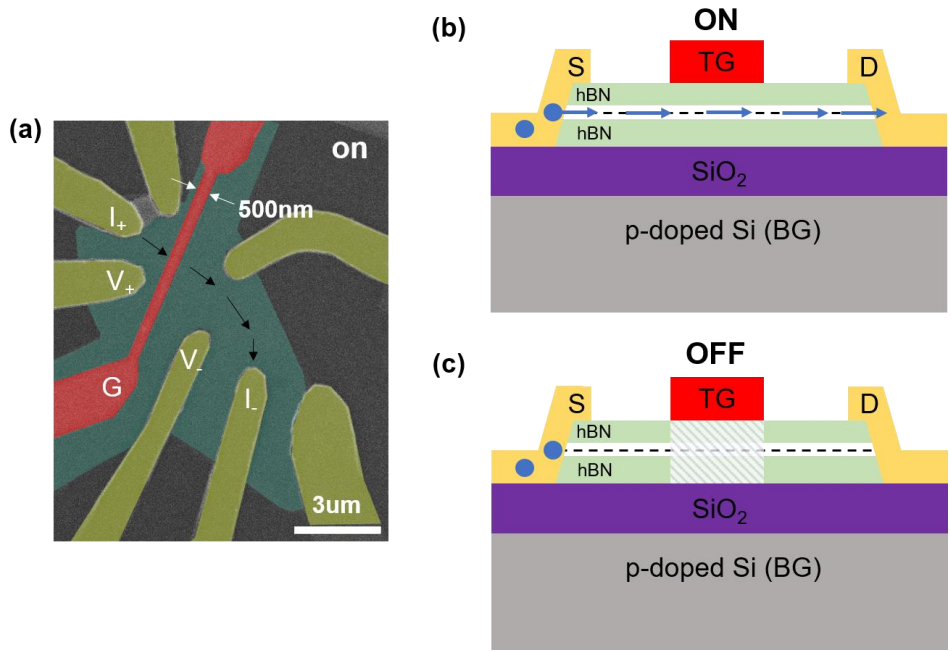

**Fig. S3** (a) Schematic at on state with the SEM image. The red and yellow regions represent the submicron top gate and electrodes for ohmic contact, respectively. The black arrows indicate one of possible current paths, whose conduction is controlled by the submicron top gate. (b), (c) Side views of the "on" and "off" states in an hBN-graphene aligned device.

Our devices were fabricated with a large bottom gate and a submicron metallic top gate, 500 nm in width, enabling precise doping control in each region. As shown in **Fig. S3(b)**, the current can flow on this device when it is in the "ON" state, by tuning the submicron top gate (G); conversely, the current cannot flow when the top-gated region is configured to the moirè-induced insulating states ("OFF" state), as depicted in **Fig. S3(c)**.

## II. Additional device properties: misalignment angle according to positions

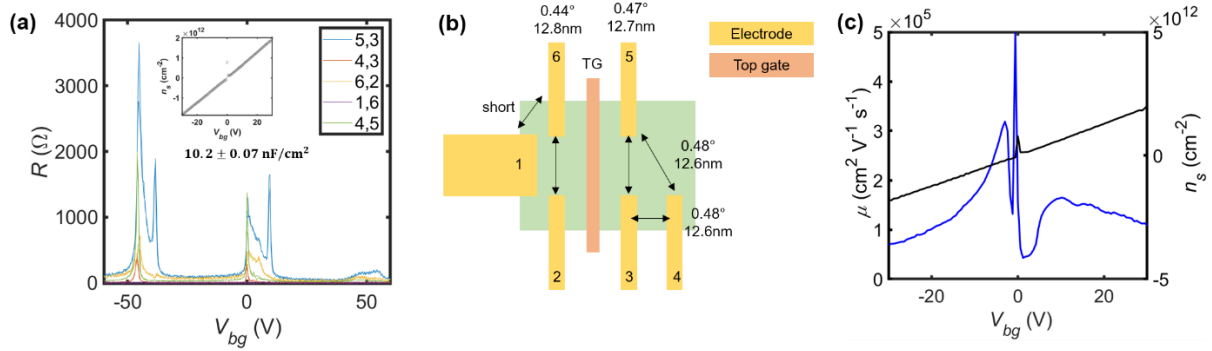

**Fig. S4** (a) The resistance according to the bottom gate. The inset graph depicts the charge density with respect to the bottom gate. These charge density values are deduced from the slope of the Hall resistance versus magnetic field graph. The slope of the inset graph represents the capacitance of the bottom insulating layer. (b) The schematic of the device and the twisted angles between electrodes. The electrode numbering corresponds to the measurement positions at (a). The average misaligned angle and superlattice parameters are determined to be 0.47° and 12.7 nm. (c) The mobility values of device A according to the bottom gate.

The misalignment angle between the hBN and graphene is determined through a transport measurement method. As the graphene possesses four degenerate states, it needs four electrons (holes) per superlattice unit cell, achieving full filling of the superlattice, $n = 4n_0$. The lattice constant of the superlattice is deduced from $n_0$ ($1/n_0 = \sqrt{3}\lambda^2/2$). Accordingly, the lattice constant of the superlattice is determined for each device region, with these values derived from **Fig. S4(a)**. Notably, the positions of FDP and h-SDP remain consistent across all configurations, indicative of the uniformity of our device. Furthermore, to accurately determine the charge density, the capacitance of the insulating layer is considered.

This capacitance is evaluated using a conventional Hall measurement technique, yielding a value of $10.2 \pm 0.07\ nF/cm^2$. By using this value and **Fig. S4(a),** the misaligned angles at each device region are computed. The average misalignment angle and superlattice constant are determined to be 0.47° and 12.7 nm. These angles exhibit consistency across all configurations, affirming the uniform fabrication of our device. We also measured the mobility by using the Hall measurement information. The full mobility value appeared at **Fig. S4(c)**.
19

## III. Additional device properties: The band gap at global and local region

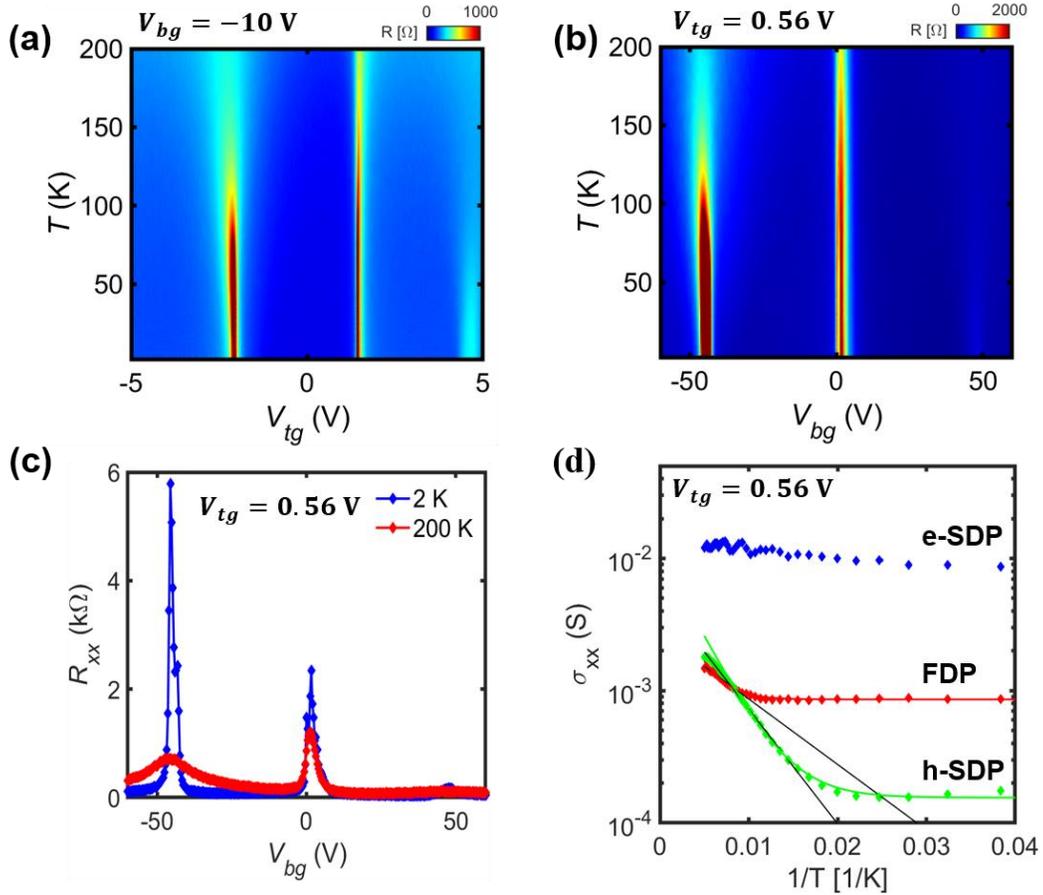

**Fig. S5** (a), (b) The resistance color plot according to top (bottom) gate and temperature. The band gaps, h-SDP, FDP and e-SDP are observed with top gate at (a) and bottom gate at (b). To avoid additional doping effects across different regions, the bottom gate is fixed with -10 V (a) and top gate is fixed with 0.56 V (b). (c) $R_{xx}$ resistance changes according to the bottom gate. (d) The ratio of conductance changes according to temperature. The h-SDP and e-SDP graph show an insulating property.

We measured the resistance variations according to the top (bottom) gate and temperature at **Fig. S5(a)** and **Fig. S5(b)**. The band gaps within the top-gated region were determined by vertical line cuts of **Fig. S5(a)**. These temperature-dependent graphs are useful to identify the optimal position corresponding to the insulating state. We measured the band gaps within the bulk region from **Fig. S5(b).** According to Fig. S5(b), the FDP, h-SDP and e-SDP positions are 0, -45.6 and 47.6 V. The h-SDP and FDP exhibit insulating behavior, as shown in **Fig. S5(d).** Through fitting with Arrhenius formula, $\sigma \propto \exp\exp\left(-\frac{\Delta}{2k_BT}\right)$, the band gaps within the global region at the FDP and h-SDP are determined to be 21 and 34 meV, respectively.

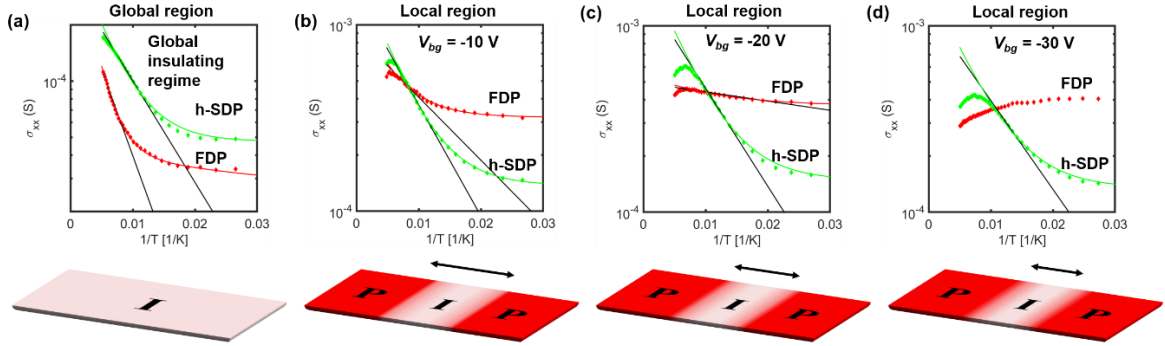

**Fig. S6** (a) The reciprocal temperature versus conductivity graph of device D at global region. The green and red dots are h-SDP and FDP. The black solid lines are fitted with the Arrhenius formula. The green and red solid curves are fitted with the combination of the Arrhenius and variable-range hopping formulas. (b), (c), (d) the reciprocal temperature versus conductivity graph of the device D at the local region. The fixed bottom gates are -10, -20, and -30 V at each graph. The schematics illustrate the variation of the insulating region (G).

During the band gap measurement, an additional tuning parameter can be controlled, i.e., the doping levels in region S-D through the bottom gate, while maintaining the insulating local region (G). As shown in **Fig. S6**, the measured band gaps diminish as the doping level of region S-D increases. The band gaps of FDP and h-SDP are 37 and 22 meV in the global insulating regime, respectively. When S and D regions are doped to be metallic, the band gaps of FDP and h-SDP are measured to 13 and 24 meV at -10 $V_{bg}$, 2 and 21 meV at -20 $V_{bg}$, no gap and 19 meV at -30 $V_{bg}$, respectively. It is unclear to us, at this point, why the measured gap sizes depend on the doping levels of region S and D, but we conjecture that the decreased bandgaps may be due to the reduced charge puddle energy $E_c$[26] as the insulated region (G) becomes narrower, as illustrated by **Fig. S6**. These data need additional theoretical investigation and simulations.

## IV. Data about other devices

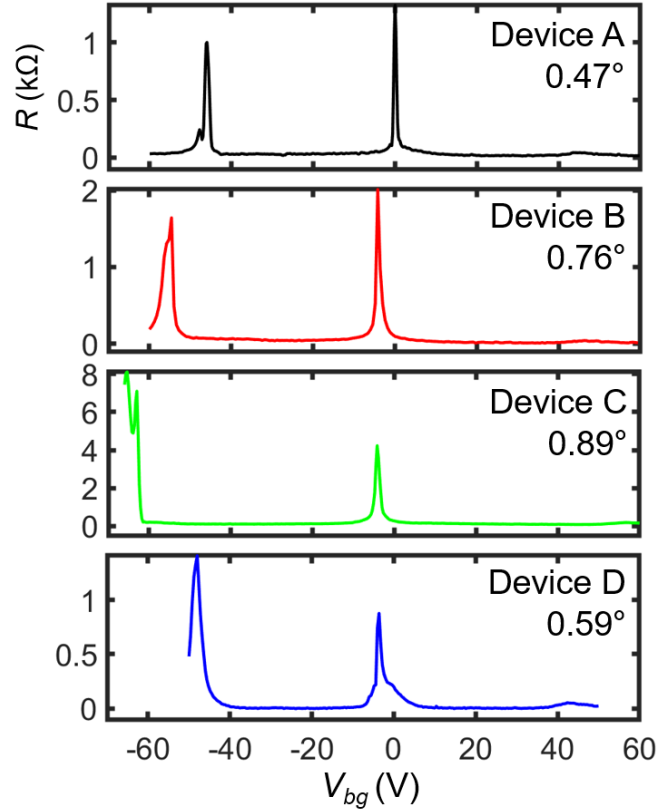

**Fig. S7** (a) Bottom gate sweep according to samples. Each sample exhibits distinctive misalignment features, evidenced by h-SDP and e-SDP. Due to the disparate positions of h-SDP and e-SDP, the misalignment angles of these devices are measured.

When the graphene is aligned with hBN sheets, an additional band gap is generated at the hole doping side. This phenomenon serves as a confirmation of the alignment status between the graphene and hBN layers within the device. Devices A, B, C and D exhibit the h-SDP, each aligned at different angles. The angles of devices A, B, C and D are 0.47°, 0.76°, 0.89° and 0.59°. As the misalignment angle approaches zero, the band gap increases.[17,18] To utilize our device effectively as a FET, we measured the device A (0.47°).

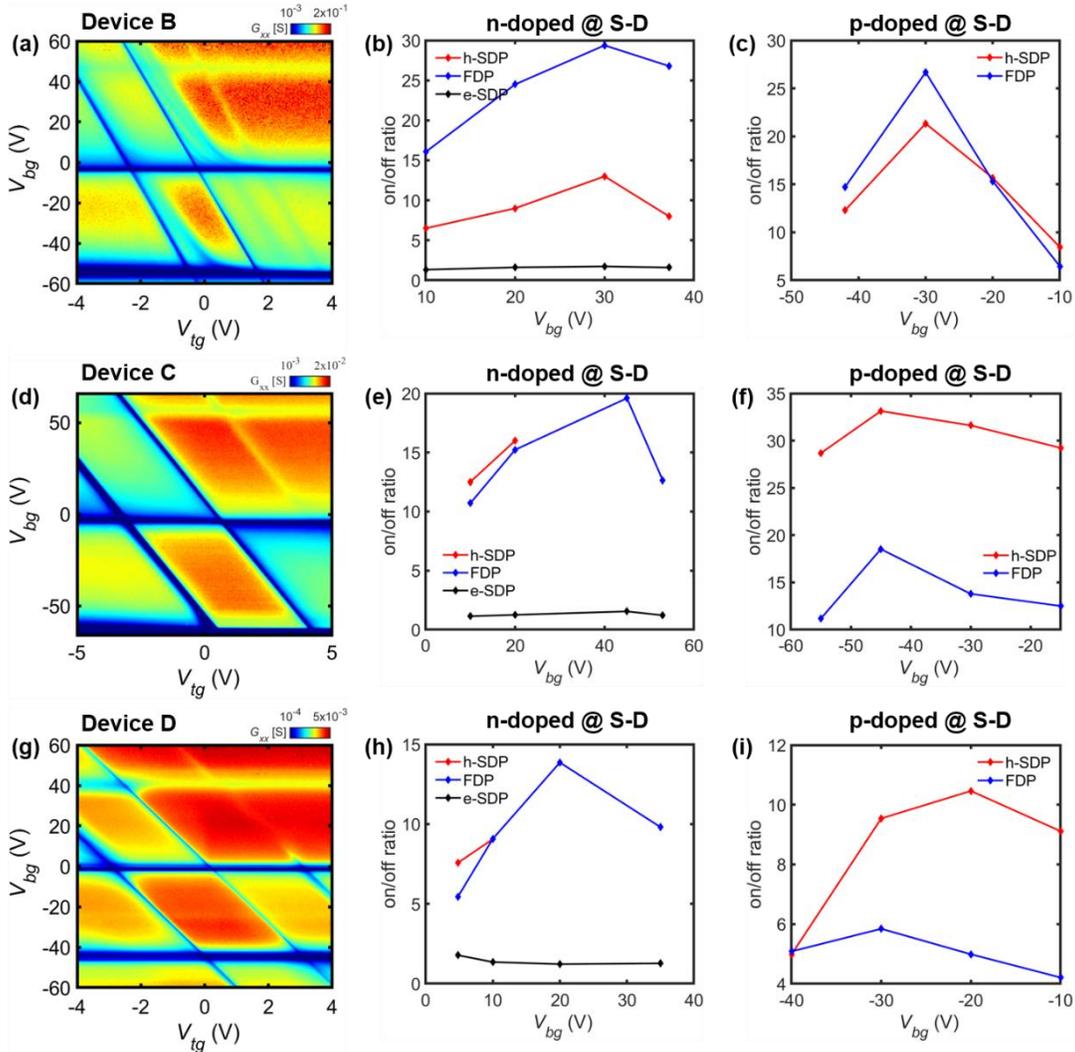

**Fig. S8** (a), (d), (g) Top and bottom gate sweep graph at device B, C and D. The misaligned angles are 0.76°, 0.89° and 0.59° at device B, C, and D. The right two panels next to the top and bottom gate graph are an on/off ratio according to doping level in the global region.

In our measurement, the other three devices, B, C, and D, showed slightly lower on/off ratios than device A. As shown in **Fig. R4**, the maximum on/off ratios for devices B, C, and D were about 30, 35, and 15, respectively, and these variations seem to come from the differences in device qualities and alignment angles. However, the qualitative behavior of the on/off ratios as a function of gate voltages were consistent throughout the devices, showing the increase in the on/off ratio at the optimal doping levels of the drain and source regions, as found in device A of the main text. The most conductive region was also observed in the *nnn* doping configuration, which was consistent with device A and can be explained by the particle-hole asymmetric moirè band structure as discussed in the main text.

## V. Comparing our devices with other high-mobility graphene devices

We compared our device to high-mobility graphene devices with a local gate electrode in 1) unaligned monolayer and 2) intrinsic Bernal bilayer graphene systems.

1) There are no band gaps in the unaligned graphene-hBN structure, which means that the on/off ratios are not able to be defined in principle. However, we note an interesting example in a ref, (*Mirza M. Elahi et al, Appl. Phys. Lett.* **114***, 013507 (2019)*), which uses ballistic electrons in a high quality monolayer graphene that realize an electron-optics system and divert the path of electrons when they control the "ON" and "OFF" states. In this case, the on/off ratio achieved was about 100, which is comparable in size to the maximum on/off ratio of our devices, ~50.

2) In Bernal bilayer graphenes (unaligned): There is no reference with a submicron top gate on a high mobility Bernal bilayer graphene that can be directly compared to ours, due to the technical difficulty of controlling the vertical electrical field to open a gap and tuning the carrier density at the same time. Instead, we found two references reporting a Bernal bilayer with a um-sized top gate with $Al_2O_3$ and $Y_2O_3$ oxidation layers (*Szafranek et al, Nano Lett.* **12***, 1324 (2012)*, ref [5], and *Kanayama et al, Sci. Rep.* **5***, 15789 (2015)*). The ref [5] was measured at room temperature with 35 maximum on/off ratio, and "Kanayama et al (2015)" was measured at 20 K with $3\times10^3$ maximum on/off ratio. However, the devices are of lower quality and mobility (the mobilities of these references are ~1,500 $cm^2 \cdot V^{-1} \cdot s^{-1}$).

We would like to mention that, while the on/off ratio, ~50, of our device is not higher than other realizations of graphene FET in these references, the focus of our work is to investigate a monolayer graphene system (the simplest structure of graphene) in a zero-angle-twisted hBN-aligned heterostructure (the most *stable* form among "twisted"-layered systems) with moirè-induce gaps and a submicron local gate. In that regard, we believe that our graphene-hBN aligned devices demonstrated the prototypical behavior of "twistronic" devices with high mobility and good reproducibility, even after the lengthy nanofabrication process including a 350 °C annealing step. Our work hopefully becomes a step forward in realizing FETs of other twisted graphenes, such as twisted bilayer or trilayer graphenes in the future.

## VI. Simulation of conductance according to doping level within region A and B

Here we provide the methodology employed for simulating quantum transport in an hBN-aligned monolayer graphene FET. The simulations were carried out using the Kwant software[27], and the system model and parameters were adapted from references.[13,28,37]

Starting from the Dirac Hamiltonian of graphene: $H_0 = vp \cdot \sigma + V(r)$, where $v = 10^6 m/s$ is the Fermi Velocity. For computational reduction, we used scalable tight binding model and set $a = a_0, t = t_0/s$, where $a_0 = 0.246nm, t_0 = 2.8eV$ are the tight binding parameter for a standard graphene and $s = 4$ is a scaling factor. Superlattice constant and orientation angle is $\lambda = \frac{a}{\sqrt{\theta^2+\epsilon^2}}$, $\phi = arctan\left(\frac{-sin\theta}{1+\epsilon-cos\theta}\right)$ for arbitrary alignment angle $\theta$, and misalignment $\epsilon = 0.0181$. The reciprocal vectors are $b_m = \hat{R}\left(\frac{2\pi m}{6}\right)b_0$, $b_0 = \hat{R}(\phi)(0,b)$ for $b = \frac{4\pi}{\sqrt{3}a}\sqrt{\theta^2+\epsilon^2}$ and rotation matrix $\hat{R}(\phi)$.

The moiré effect from h-BN adds superlattice potentials on the Dirac Hamiltonian in two terms, onsite potential and hopping amplitudes. The onsite potential terms for sublattice A and B is

$$V_A(r) = \epsilon^+ + \epsilon^- + V(r), V_B(r) = \epsilon^+ + \epsilon^- + V(r)$$

where

$$\epsilon^+ = E_0[cos(g_1 r) + cos(g_1 r) + cos(g_1 r)], \quad \epsilon^- = E_0[sin(g_1 r) + sin(g_1 r) + sin(g_1 r)]$$

($E_0 = 0.05\hbar bv(1 + \theta^2/\epsilon^2)^{-0.5}$ is moiré energy parameters) are moiré potential and

$$V(x,y) =$$

$$E_1 \quad \left(x < -\frac{L_m}{2} + \lambda I_L(y) - \frac{w_j}{2} \text{ or } x > \frac{L_m}{2} + \lambda I_R(y) + \frac{w_j}{2}\right)$$

$$E_2 \quad \left(-\frac{L_m}{2} + \lambda I_L(y) + \frac{w_j}{2} < x < \frac{L_m}{2} + \lambda I_R(y) - \frac{w_j}{2}\right)$$

$$\frac{E_1 + E_2}{2} - \frac{(E_1 - E_2)}{2w_j}(x + L_m) + \delta \quad \left(-\frac{L_m}{2} + \lambda I_L(y) - \frac{w_j}{2} < x < -\frac{L_m}{2} + \lambda I_L(y) + \frac{w_j}{2}\right)$$

$$\frac{E_1 + E_2}{2} + \frac{(E_1 - E_2)}{2w_j}(x - L_m) + \delta \quad \left(\frac{L_m}{2} + \lambda I_R(y) - \frac{w_j}{2} < x < \frac{L_m}{2} + \lambda I_R(y) + \frac{w_j}{2}\right)$$

is electrostatic potential induced considering random potential $\delta$, rough boundary $\lambda I_L(y)$, $\lambda I_R(y)$ and junction depth $w_j$ for each junction.

Lastly, we replaced the hopping term to

$$t_{i,j} = t + \frac{2}{3}A_x(r_{ij})sin(\varphi_{ij}) - \frac{2}{3}A_y(r_{ij})cos(\varphi_{ij})$$

where $A = (A_x, A_y) = E_0 \frac{\epsilon}{b\sqrt{\theta^2 + \epsilon^2}}\left(-\frac{\partial f_-}{\partial y}, \frac{\partial f_-}{\partial x}\right)$, $f_- = sin(g_1 r) + sin(g_1 r) + sin(g_1 r)$, $r_{ij}$ as position of middle of the hopping bond and $\varphi_{ij}$ is angle between displacement of hopping and x-axis.

The simulation was done using width $W = 400\ nm$, length $L = 1200\ nm$(400nm for each region). Conversing the result of the simulation to top/bottom gate voltage following formula.

$$V_{bg} = \frac{e}{\pi C_b}\left(\frac{E_1}{\hbar v_F}\right)^2 sgn(E_1) \qquad V_{tg} = \frac{e}{\pi C_t}\left[\left(\frac{E_2}{\hbar v_F}\right)^2 sgn(E_2) - \left(\frac{E_1}{\hbar v_F}\right)^2 sgn(E_1)\right]$$

# References

[35] Y. Kim, P. Herlinger, T. Taniguchi, K. Watanabe, and J.H. Smet, "Reliable Postprocessing Improvement of van der Waals Heterostructures," ACS Nano **13**(12), 14182–14190 (2019).

[36] L. Wang, I. Meric, P.Y. Huang, Q. Gao, Y. Gao, H. Tran, T. Taniguchi, K. Watanabe, L.M. Campos, D.A. Muller, J. Guo, P. Kim, J. Hone, K.L. Shepard, and C.R. Dean, "One-Dimensional Electrical Contact to a Two-Dimensional Material," Science **342**(6158), 614–617 (2013).

[37] S.W. LaGasse, and J.U. Lee, "Theory of Landau level mixing in heavily graded graphene p-n junctions," Phys. Rev. B **94**(16), 165312 (2016).